\documentclass[fleqn,10pt]{wlscirep}
\usepackage[utf8]{inputenc}
\usepackage[T1]{fontenc}
\title{Isotope-selective Ion Trapping via Sympathetic Cooling using a Surface-Electrode Trap with a Hole for Collimated Atomic Loading}
\usepackage{subcaption} 
\author[1,*]{Masanari Miyamoto}
\author[2,+]{Takashi Higuchi}
\author[3,+]{Kentaro Furusawa}
\author[3,+]{Norihiko Sekine}
\author[3,1,+]{Kazuhiro Hayasaka}
\author[1,3,4,+]{Utako Tanaka}
\affil[1]{Graduate School of Engineering Science, Osaka University}
\affil[2]{Institute for Integrated Radiation and Nuclear Science, Kyoto University}
\affil[3]{National Institute of Information and Communications Technology}
\affil[4]{Center of Quantum Information and Quantum Biology, Osaka University}

\affil[*]{u957000h@ecs.osaka-u.ac.jp}

\affil[+]{these authors contributed equally to this work}


\begin{abstract}
We developed a surface-electrode ion trap with a square hole measuring $40\,\mathrm{\mu m}$ for atomic loading. The hole was fabricated using anisotropic etching of a silicon substrate and was designed to minimize potential distortion in the trapping region. By introducing the atomic beam through the hole, we achieved enhanced isotope selectivity and experimentally demonstrated the selective trapping of calcium isotope ions using an atomic oven. We successfully prepared isotope ion pairs directly from the oven via sympathetic cooling at a rate comparable to that achieved using ablation loading. The sympathetic cooling process occurred on the order of a few seconds. We demonstrated the direct generation of an ion chain above the through-hole. This approach can be applied for trapping a wide range of ion species using a remarkably simple experimental setup, making it desirable for several applications such as quantum-charge-coupled-device (QCCD) architectures and precision measurements of isotope shifts.
\end{abstract}
\begin{document}

\flushbottom
\maketitle
%
%
\thispagestyle{empty}

\section*{Introduction}

Surface-electrode ion traps are anticipated for the practical implementation of ion-trap quantum computers, particularly in the realization of a quantum-charge-coupled-device (QCCD) architecture \cite{Wineland1998, Kielpinski2002, QCCD}, because they enable realization of a variety of electrode layouts. One of the issues of surface-electrode ion traps is that atomic deposition on the electrodes induces patch potentials, generating electric-field noise that leads to ion heating \cite{ion_heating}. This phenomenon presents a significant challenge for quantum computing and precision measurements. Furthermore, exposure to the atomic beam causes time-dependent degradation of the surface, leading to distortion of trapping potential and ion instability \cite{PhysRevA.65.063407}. Therefore, suppressing trap surface contamination and mitigating time-dependent degradation induced by atomic deposition are significant objectives.

\par
One approach to suppress surface contamination and time-dependent degradation involves placing the atomic source behind the trap and loading ions through a small aperture, a strategy proposed and demonstrated for surface-electrode ion traps \cite{Charles_Doret_2012, PhysRevX.13.041052}. In previous work \cite{PhysRevX.13.041052}, a 2D magneto-optical trap (MOT) was used as a source of neutral atoms, and the loading time was made to be as fast as milliseconds for one ion. Such a loading method using MOT is sophisticated and requires several laser light sources and relatively large vacuum systems.

\par
Another crucial aspect for realizing a QCCD architecture is the treatment of ion heating induced by transports. To avoid perturbing the internal states of ions used as qubits, a cooling mechanism that it does not rely on the internal transitions. Sympathetic cooling is well suited to this purpose because it does not involve qubit transition and has been successfully employed in various fields such as quantum operations and optical clocks \cite{long_coherence, 24_ion_entanglement, Al_quantum_clock}. These research efforts require specific isotope species, such as $^{171}\mathrm{Yb}^+$ and $^{27}\mathrm{Al}^+$. Therefore, isotope selectivity is a critical aspect of ion loading.
\par
In this study, we developed a surface-electrode trap featuring a 40-$\,\mathrm{\mu m}$ square through-hole. Surface-electrode traps with through-holes are generally used to prevent contamination of the trap surface caused by atomic deposition by introducing the atomic beam from the back side. The loading scheme can be exploited to enhance the isotope selectivity of resonant photoionization by illuminating the vertically propagating atomic beam with a horizontally directed ionization laser. We experimentally verified high isotope selectivity and enabled selective loading of different ion species. The hole was fabricated by anisotropic etching to create a structure that suppresses the electric field noise that arises from the build-up of charge on the sidewalls of the hole \cite{8116614, Jung_2021}. 

\par
An atomic oven, used as a source, is placed on the backside of the electrode behind the hole, thereby minimizing atomic contamination on the electrode surface. Atoms are loaded through this hole, ionized and cooled by laser beams covering the entire beam size. This loading method suppresses contamination and achieves a high selectivity of ion species to be loaded. The initial loading of ions can be performed via sympathetic cooling with the trap, which significantly reduces the required experimental resources. The initial loading via sympathetic cooling using an ablation laser and blade-type traps has been reported \cite{Sympathetic_cooling_hot_ion, SunChenglong2022El4i}. By contrast, we performed the initial loading via sympathetic cooling with an atomic oven and a surface-electrode trap. Although loading with a resistive oven generally requires more time than loading with an ablation laser, we find that we could load pairs of ions, with $^{40}\mathrm{Ca}^+$ ($^{44}\mathrm{Ca}^+$) as the coolant ions and $^{44}\mathrm{Ca}^+$ ($^{40}\mathrm{Ca}^+$) as the corresponding sympathetically cooled ions, in a comparable time with the ablation method \cite{SunChenglong2022El4i}. This approach allows for simple and compact experimental apparatus, desirable for various applications such as QCCD architectures and precision measurements of isotope shifts\cite{Knollmann2019}.

\section*{Experimental apparatus}
\begin{figure}[htbp]
  \centering
  \includegraphics[width=15cm]{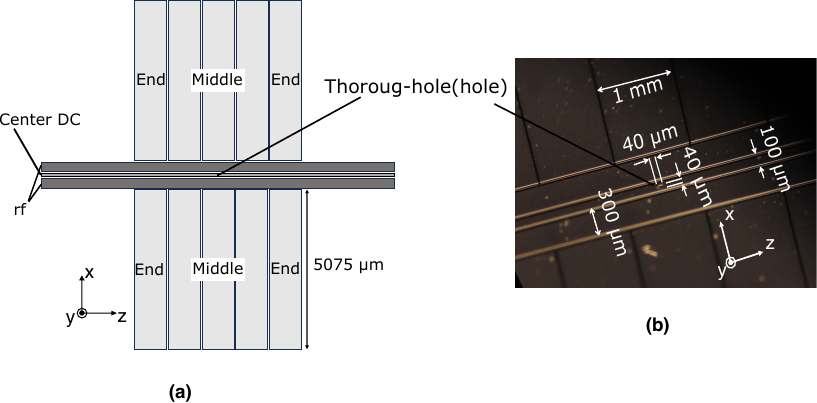}
  \caption{Overall structure of the trap. (a) Electrode arrangement and sizes are shown. Two RF electrodes, four End DC electrodes, six Middle DC electrodes, and one Center DC electrode are symmetrically arranged. (b)Image of the through-hole. The hole is located at the center of the trap and has a square shape with a side length of 40 $\mathrm{\,\mu m}$. The Ca beam is introduced through this hole.}\label{fig:Figure1}
\end{figure}
In this section, we detail the trap structure and experimental setup. The trap consists of two RF and eleven DC electrodes. The DC electrodes include four End, six Middle and one Center electrodes, named after their locations in the trap layout (Fig.~\ref{fig:Figure1}(a)). A 40-$\mathrm{\mu m}$ square hole for atom introduction is fabricated at the center of the Center electrode (Fig.~\ref{fig:Figure1}(b)). All electrodes are gold plated and adhere to the Si substrate. The calcium oven is positioned 3 mm below the trap. The calcium source consists of a stainless steel tube with an outer diameter of 1.26 mm and an inner diameter of 0.9 mm. One end of the tube is sealed by threading a screw into it, and the opposite open end serves as the exit for the atomic beam. The screw overlaps with the tube by 2 mm, forming the base of the oven. The screw acts as the ground electrode, and a wire is welded to the opposite end of the stainless steel tube to apply a current ranging from 1.5--3.6 A. The estimated oven temperature, inferred from vacuum conditions, is in the range of 480--550 K.

\subsection*{Shaped through-hole}\label{subsec_shaped_throughhole}
The hole placed at the center electrode had two functions: to introduce an atomic beam from the backside of the trap electrode to prevent accumulation of contamination on the electrode surface, and to collimate the atomic beam for obtaining high-resolution spectra in the photo-ionization transition. When determining the size of the through-hole, we performed numerical calculations of the trapping potential with the existence of a hole at the center electrode. We did not find any substantial potential distortion at the ion height of about $200\,\mathrm{\mu m}$ above the electrode surface, provided that the hole size was less than 60-$\mathrm{\mu m}$ square, for which the distortion was less than no greater than $10^{-3}$ V. On the other hand, if the size was too small, atoms cannot be efficiently loaded. Finally, we decided the hole size to be $40\,\mathrm{\mu m}$ square.

\par
Through the hole, the atomic beam is collimated, which allows us to obtain good isotope selectivity. Owing to the geometric relationship between the oven and through-hole, the atomic beam is collimated and expected to have a divergence angle of approximately 0.76$^{\circ}$.

\par
To avoid the effects of charging on the substrates, the trap structure must be designed such that the trapped ions are not exposed to the insulators \cite{Jung_2021}. Therefore, we designed a through-hole that has a square pyramid shape underneath the trap electrode. Owing to the characteristics of Si crystals, a $54.7^\circ$-sloped hole can be realized in a (100)-oriented Si substrate via anisotropic etching of the silicon substrate.

\par
\begin{figure}[tb]
  \centering
  \includegraphics[width=15cm]{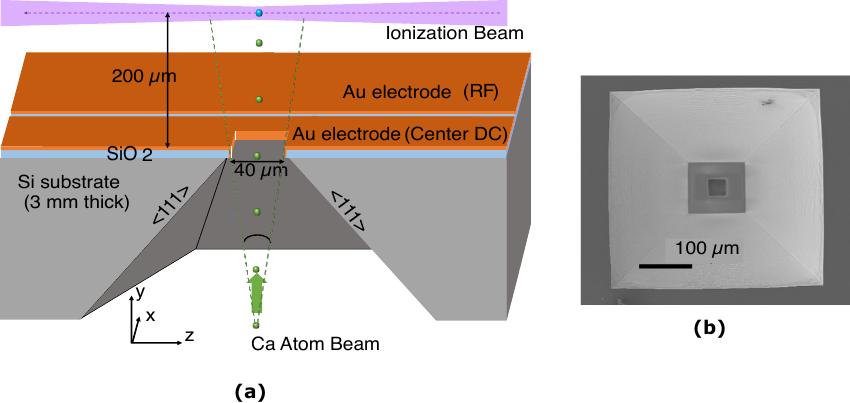}
  \caption{(a)Design of the through-hole. By using anisotropic etching, Si base forms a square pyramid shape. After collimated by these structure, the calcium beam crosses with ionization beams which propagate parallel to trap surface. (b)SEM image of the hole taken from backside.}\label{fig:Figure2}
\end{figure}

The trap was fabricated at the National Institute of Communications and Technology, Tokyo, Japan. The structure of the through-hole is shown in Fig.~\ref{fig:Figure2}(a). For the trap, a 22-$\mathrm{\mu m}$-thick SiO$_2$ layer was fabricated on the top of a silicon substrate using the direct bonding technique to ensure a flat surface. The surface electrodes were patterned using photolithography and electroplating. Anisotropic etching was then performed from the backside of the substrate using TMAH solution (25\%) with alkaline-resistant photoresist as a mask material. The SiO$_2$ layer serves served as an etch stop. Finally, a hole was made in the SiO$_2$ layer using the focused ion beam~(FIB) method. Figure~\ref{fig:Figure2}(b) shows a scanning electron microscope~(SEM) image of the hole taken from the backside of the trap electrode.

\subsection*{Experimental setup}\label{subsec_experimental setup}
\begin{figure}[tb]
  \centering
  \includegraphics[width=15cm]{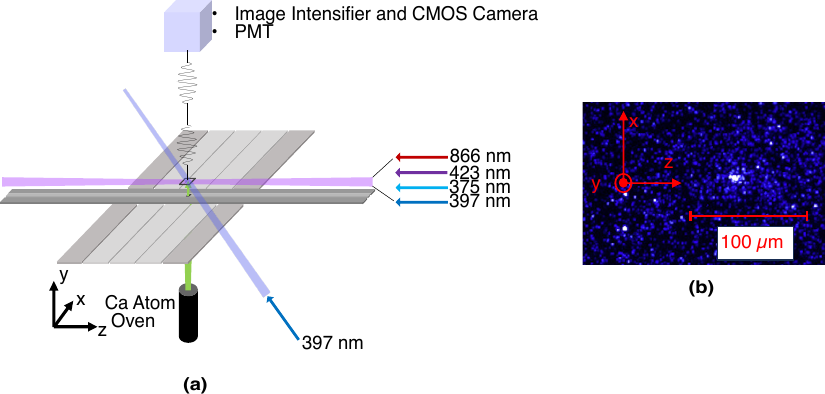}
  \caption{(a)Schematic of the ion fluorescence detection system and laser setup. Ions introduced into the trap are cooled by the 397-$\rm{nm\,}$ and 866-$\rm{nm\,}$ lasers incident along the z-axis, as well as by a 397-$\rm{nm\,}$ laser incident at a 45$^{\circ}$ angle to the z-axis. The 397-$\rm{nm\,}$ lasers are split into two paths by a half mirror to irradiate from two directions.
 (b) Image of an ion trapped above the through-hole. 
}\label{fig:Figure3}
\end{figure}

The experimental setup consists of four laser sources: 423-$\rm{nm}$ external cavity diode laser (ECDL), free-running 375-$\rm{nm}$ laser, 397-$\rm{nm}$ ECDL, and 866-$\rm{nm}$ ECDL (Fig.~\ref{fig:Figure3}(a)). As described in Section 2, calcium atoms are ionized by the intersection of the 423-$\rm{nm}$ and 375-$\rm{nm}$ lasers above the through-hole. 

Once the calcium ions are loaded, they are cooled via Doppler cooling. The 397-$\rm{nm}$ laser drives the $^2S_{1/2} \longrightarrow\,^2P_{1/2}$ transition, and the 866-$\rm{nm}$ laser serves as a repumper for the $^2D_{3/2} \longrightarrow\,^2P_{1/2}$ transition. During this process, all laser beams overlap and propagate along the z-axis. The 397-$\rm{nm}$ laser is introduced at a 45$^\circ$ angle to the z-axis using a half-mirror.

For observation, ion images in the trap are captured using an image intensifier and a CMOS camera (Fig.~\ref{fig:Figure3}(b)). The fluorescence intensity is measured with a photomultiplier tube (PMT)
(Fig.~\ref{fig:Figure3}(a)).

\section*{Experimental results}
We trapped $^{44}\rm{Ca}^+$ and $^{40}\rm{Ca}^+$ ions selectively using naturally occurring calcium isotopes. The natural abundances of $^{40}\rm{Ca}$ and $^{44}\rm{Ca}$ are 96.9$\%$ and 2.4$\%$, respectively. We then performed initial loading via sympathetic cooling. In our experiments, either of the two isotopes ($^{44}\rm{Ca}^+$ or $^{40}\rm{Ca}^+$) could serve as the coolant and sympathetically cooled ion.

\subsection*{Isotope selective trap}\label{subsec_cross_beam_spectroscopy}
\begin{figure}[htbp]
  \centering
  \includegraphics[width=15cm]{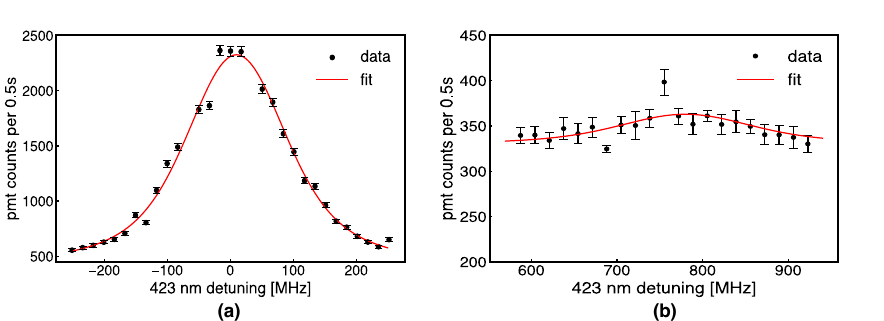}
  \caption{Cross-beam spectroscopy. 
  (a) Measurement results of the $^{40}\mathrm{Ca}^+$ spectrum. The red curve represents the best-fit Voigt function. 
  (b) Measurement results of the $^{44}\mathrm{Ca}^+$ spectrum. The red curve shows a Voigt function fitted with physically reasonable parameter constraints. The isotope shift of $^{44}\mathrm{Ca}^+$ is approximately 757\,MHz.
  }\label{fig:Figure4}
\end{figure}

\par
We measured the cross-beam spectrum of a calcium atomic beam using a 423-nm laser to investigate isotope selectivity. Based on the isotope shift values reported in \cite{Isotope-selectivity}, we scanned the ionization laser from detunings of $-250$\,MHz to $+950$\,MHz, taking the resonance of $^{40}\mathrm{Ca}^+$ as the zero point. The resulting spectra near the $^{40}\mathrm{Ca}^+$ and $^{44}\mathrm{Ca}^+$ resonances are shown in Fig.~\ref{fig:Figure4}. To analyze these data, we employed the Voigt function, which is the convolution of Lorentzian and Gaussian functions. The Gaussian component corresponds to Doppler broadening caused by the horizontal velocity of the atomic beam, enabling us to deduce the divergence angle of the beam. The Lorentzian component includes the natural linewidth of calcium (approximately 35.4\,MHz \cite{NIST_data_base}), as well as power broadening and transit-time broadening, both estimated from our experimental conditions \cite{AtomicPhysics}. The transit-time broadening depends on the length of time the atoms spend in the laser beam, which is determined by the beam velocity and diameter of the laser beam.In this experiment, the oven was supplied with a current of 3.54\,A, and according to vapor pressure curve \cite{honig1969vapor}, corresponding to a temperature of 530\,K. Under these conditions, the most probable velocity of the atoms is roughly 573\,m/s. The ionization laser used in the measurement had an intensity of 50\,$\rm{\mu}$W with a beam diameter of 250\,$\mu$m, leading to a saturation intensity of $6.11\times10^2\,\mathrm{W/m^2}$. From these parameters, we estimated the relative contributions of power broadening and transit-time broadening and then fitted the observed spectrum with the Voigt function. The Lorentzian and Gaussian widths were 60.1(1.17)\,MHz and 50.2(1.00)\,MHz, respectively, yielding a beam divergence angle of 2.54(0.05)$^\circ$. This angle is greater than the predicted value ($0.76^{\circ}$) corresponding to the divergence angle from the center of the oven. The angle obtained from the measurement data is considerably larger than the predicted value, likely because atoms are emitted not only from the center but also from other regions of the oven. Nevertheless, based on the measured angle, the calcium atomic beam is emitted radially from an atomic source with a diameter of 93\,µm.

The atomic beam was confirmed to be collimated. The intensity peak at 750 MHz (Fig.~\ref{fig:Figure4}(b)) was attributed to $^{44}\rm{Ca}$ because the ratio of the fluorescence peak to the natural abundance ratio is similar for Figs.~\ref{fig:Figure4}(a) and \ref{fig:Figure4}(b). The Doppler broadening was significantly smaller than the isotope shift, indicating high isotope selectivity. From these results and other isotope shift values \cite{Isotope-selectivity}, we trapped $^{40}\rm{Ca}^+$ and $^{44}\rm{Ca}^+$ selectively by appropriately detuning each laser. It took approximately 20\,s to trap $^{40}\rm{Ca}^+$ and 30\,s to trap $^{44}\rm{Ca}^+$ owing to the significant difference in their natural abundance.

\subsection*{Sympathetic cooling experiment}
\begin{figure}[htbp]
    \centering
    \includegraphics[width=15cm]{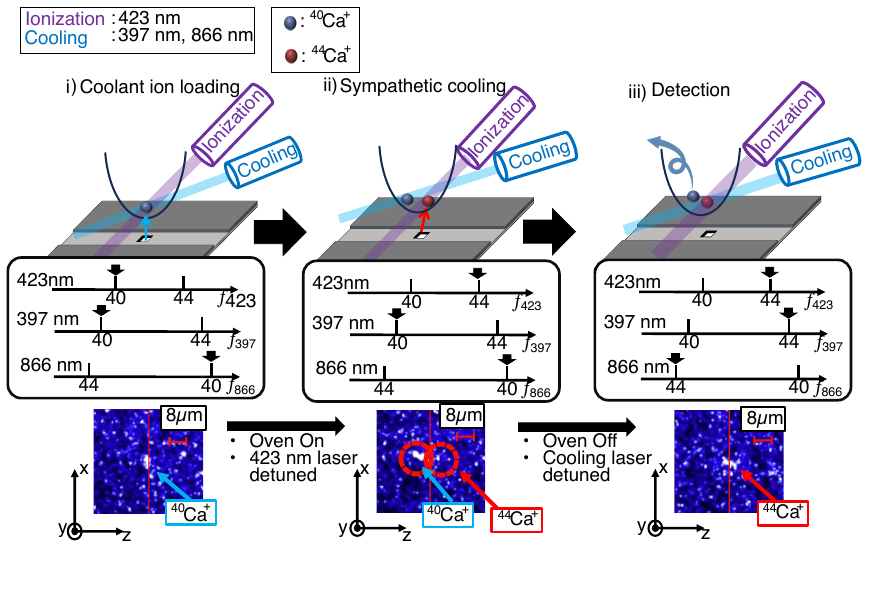}
    \caption{Experimental procedure and results of the sympathetic cooling experiment. After loading the coolant ions and trapping the hot ions through sympathetic cooling, the species of the sympathetically cooled ions was identified via Doppler cooling. The upper panels show the schematics, the middle panels show the operational settings of the lasers, and the lower panels show the ion images. Only one 397-$\rm{nm}$ and one 866-$\rm{nm}$ laser were used. During the detuning stage before detection, it was observed that when $^{40}\rm{Ca}^+$ were the coolant ions, they escaped due to laser heating, whereas when $^{44}\rm{Ca}^+$ were the coolant ions, they were not affected by laser heating but often escaped during the few seconds while the 397-$\rm{nm}$ laser frequency was far from the resonance frequency of $^{44}\rm{Ca}^+$ and $^{40}\rm{Ca}^+$ ions.}
    \label{fig:Figure5}
\end{figure}

Guggemos \textit{et al}. \cite{Sympathetic_cooling_hot_ion} trapped and detected hot ions via sympathetic cooling in a linear Paul trap. In the same manner, we trapped hot isotope ions via sympathetic cooling using selectively trapped ions. The experimental setup is simplified owing to the use of isotope ions. The experimental procedure (Fig.~\ref{fig:Figure5}) is as follows: i) We trap $^{40}\rm{Ca}^+$ as a coolant ion via Doppler cooling. ii) We detune the 423-$\rm{nm}$ laser by 750$\rm{\,MHz}$ and introduce $^{44}\rm{Ca}^+$ above the trap. iii) We detune the 397-$\rm{nm}$ laser by 842$\rm{\,MHz}$ and the 866-$\rm{nm}$ laser by $-4495\rm{\,MHz}$ to identify the sympathetically cooled ion species. The $^{44}\rm{Ca}^+$ ions introduced in Step ii) require almost no cooling because the 397-$\rm{nm}$ and 866-$\rm{nm}$ lasers are tuned for cooling the $^{40}\rm{Ca}^+$ ions; therefore, if we can trap $^{40}\rm{Ca}^+$ ions, it implies that hot $^{44}\rm{Ca}^+$ ions are trapped by sympathetic cooling. Because the surface-electrode trap has a shallow potential, we thought that the atomic beam and coolant ions would collide and the coolant ions would escape; however, we could trap hot $^{44}\rm{Ca}^+$ ions by sympathetic cooling (Fig.~\ref{fig:Figure5}).

\par
Sympathetically cooled ions do not fluoresce; therefore, their presence was inferred from the displacement of the coolant ions. However, because a natural atomic source was used, other isotopic ions could get trapped as well. Therefore, in Step iii), we identified the species of the trapped ion. In the same manner, we were able to trap hot $^{40}\rm{Ca}^+$ ions via sympathetic cooling using $^{44}\rm{Ca}^+$ as the coolant ion. In this case, $^{40}\rm{Ca}^+$ ions were trapped via sympathetic cooling and detected, even though they were subjected to laser heating owing to a small isotope shift in the red direction (842$\rm{\,MHz}$). In both cases, we conducted control experiments without preparing the coolant ions and confirmed that hot ions could be trapped via sympathetic cooling.
\par
\begin{figure}[htbp]
    \centering
    \includegraphics[width=15cm]{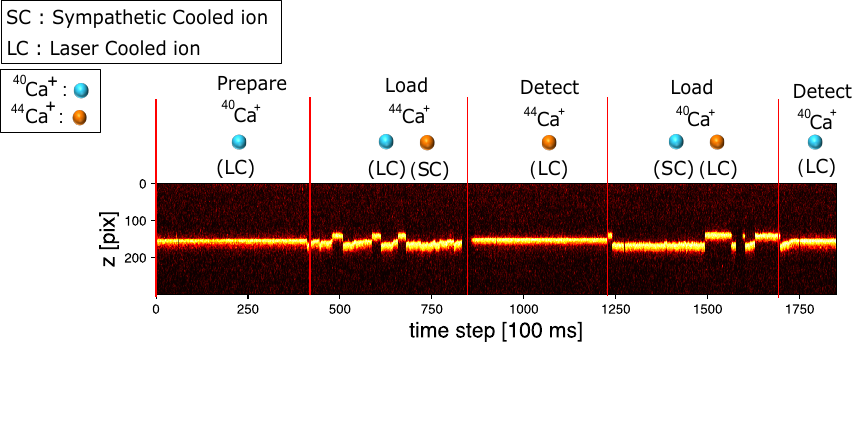}
    \caption{Time-lapse images obtained by performing the same experiment continuously. The experiment is the one shown in Fig.~\ref{fig:Figure5}, as well as the case where the roles of the coolant ion (LC) and sympathetically cooled ion (SC) are reversed. Except for the initially loaded $^{40}\mathrm{Ca}^+$ ions, all ions were loaded via sympathetic cooling. Note that the positions of $^{40}\mathrm{Ca}^+$ and $^{44}\mathrm{Ca}^+$ ions hop.}
    \label{fig:Figure6}
\end{figure}
\begin{figure}[htbp]
    \includegraphics[width=15cm]{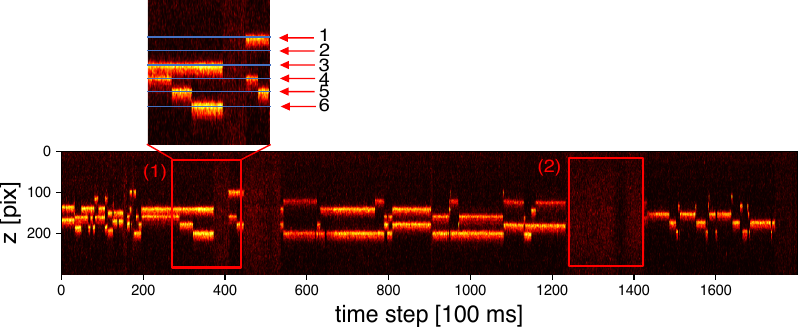}
    \caption{Extended version of the experiment shown in Fig.~\ref{fig:Figure6}, where two $^{44}\mathrm{Ca}^+$ ions were used as coolant ions (LC) and $^{40}\mathrm{Ca}^+$ ions were selectively loaded, thereby generating various isotope ion chains. Except for the initial two $^{44}\mathrm{Ca}^+$ ions, all ions were loaded via sympathetic cooling. The region labeled (1) indicates the maximum number of ions that can be crystallized. In region (2) and beyond, one of the $^{44}\mathrm{Ca}^+$ ions (LC) escaped, resulting in a state where multiple $^{40}\mathrm{Ca}^+$ ions (SC) were trapped with only a single $^{44}\mathrm{Ca}^+$ ion acting as the coolant ion.}
    \label{fig:Figure7}
\end{figure}

We performed experiments in which $^{40}\mathrm{Ca}^+$ and $^{44}\mathrm{Ca}^+$ served as the coolant ions to trap $^{44}\mathrm{Ca}^+$ and $^{40}\mathrm{Ca}^+$, respectively. The time-lapse images of the results are shown in Fig.~\ref{fig:Figure6}. We conducted an experiment to directly create isotope-ion chains on the through-hole by selectively trapping isotope ions solely via sympathetic cooling. Figure~\ref{fig:Figure7} shows the results of generating ion chains by loading $^{40}\mathrm{Ca}^+$ ions via sympathetic cooling when two $^{44}\mathrm{Ca}^+$ ions acted as coolant ions.

\par
From region (1) in Fig.~\ref{fig:Figure7}, the maximum number of $^{40}\mathrm{Ca}^+$ ions that can crystallize with two $^{44}\mathrm{Ca}^+$ ions was found to be four. In region (2), more than four $^{40}\mathrm{Ca}^+$ ions are loaded together with two $^{44}\mathrm{Ca}^+$ ions, resulting in the dominance of modes that cannot be cooled by sympathetic cooling. Consequently, the crystal melts, and one of the $^{44}\mathrm{Ca}^+$ ions eventually escapes. Mao \textit{et al}. \cite{PhysRevLett.127.143201} demonstrated that by optimizing the positions of the coolant ions, two coolant ions can efficiently cool up to fourteen sympathetically cooled ions to the Doppler limit. Therefore, precise control of the coolant ion positions could enable the loading of a longer isotope ion chain.
\par
In our experiment, no intentional operations, such as swapping ions, were performed. Nonetheless, the ion fluorescence images in Fig.~\ref{fig:Figure7} indicate that unintended hopping occurred frequently, preventing control of the coolant ion positions. Consequently, the number of ions that could be cooled was fewer than that in Ref.~\cite{PhysRevLett.127.143201}. We believe that our vacuum condition ($8.67\times 10^{-8}\,$Pa), which increases ion heating, was a major factor contributing to ion hopping. In addition, as described in Discussion and Conclusion, potential distortion caused by the through-hole may also play a role. A previous study on Paul linear traps reported that ion hopping became more pronounced when the potential anisotropy was high \cite{PhysRevA.83.063401}.

\par
In the studies by Guggemos \textit{et al}. and Sun \textit{et al}. \cite{Sympathetic_cooling_hot_ion,SunChenglong2022El4i}, atoms were introduced via an ablation laser, and owing to the large initial energy and incidence angle, the time required to cool a hot ion via sympathetic cooling was approximately 52 seconds. By contrast, by using an atomic oven, we were able to cool a hot ion via sympathetic cooling on the order of a few seconds. Although it takes about 30\,s to trap $^{44}\mathrm{Ca}^+$ ions by Doppler cooling, it requires only 30--35\,s to trap $^{44}\mathrm{Ca}^+$ ions via sympathetic cooling. Because the atoms of the oven are continuously supplied, precisely measuring the sympathetic cooling time is difficult. In any case, laser cooling allows ion loading within approximately 20 s for $^{40}\mathrm{Ca}^+$ and 30 s for $^{44}\mathrm{Ca}^+$ ions. By contrast, when using only sympathetic cooling, ions can be loaded within approximately 25 s for $^{40}\mathrm{Ca}^+$ and 35 s for $^{44}\mathrm{Ca}^+$ ions. Therefore, ion pairs can be prepared in about 1 min.
\section*{Discussion and Conclusion}\label{sec_Discussion_and_Conclusion}
In this study, we conducted experiments on selective ion generation and sympathetic cooling of isotope ions using a surface-electrode ion trap with a 40-$\mathrm{\mu m}$ square hole for atomic loading. Rotating the principal axis of the trapping potential enables efficient laser cooling using only horizontally propagating lasers \cite{IQBAL2024100208,Romaszko_2020}. In our experiment, we applied asymmetric DC voltages to rotate the principal axis. However, we found that even with symmetrically applied DC voltages, where the principal axis is theoretically not rotated, ions could still be trapped and cooled using only horizontally propagating lasers. One possible explanation for this phenomenon is that the principal axis of the potential locally rotates near the through-hole. To investigate this, we measured the ion position and trapping frequencies using forced oscillation in the vicinity of the hole. The results showed that both the ion position and trapping frequencies exhibited linear variations up to the top of the hole. However, directly above the hole, while $\omega_z$ continued to change continuously, the forced oscillation of $\omega_x$ became undetectable within a certain region. This suggests that despite our design considerations to minimize the influence of the hole on the trapping potential, the presence of the hole introduced potential disturbances. These disturbances likely caused a localized rotation of the principal axis. Iqbal and Nizamani \textit{et al}. \cite{IQBAL2024100208} reported that if the principal axis rotates by more than 6 degrees within the trapping region, laser cooling can still be sufficiently effective. Our numerical calculations confirmed that, in the absence of hole-induced local axis rotation, an asymmetric DC voltage application could tilt the principal axis by more than 6 degrees, making stable ion trapping feasible even if the hole did not induce local rotation.

\par
We investigated the selective trapping of isotope ions and found that collimating the atomic beam through the through-hole effectively suppressed Doppler broadening. The fluorescence spectrum shown in Fig.~\ref{fig:Figure4} confirms high isotope selectivity. In an experiment using a conventional horizontal atomic oven without an aperture, where a 423-nm ionization laser was tuned to $^{44}\mathrm{Ca}^+$ resonance, we frequently observed $^{40}\mathrm{Ca}^+$ ions getting trapped instead. By contrast, when using the through-hole, only about 14\% of the cases resulted in $^{40}\mathrm{Ca}^+$ ions getting trapped first. This was attributed to the slight overlap between the tail of the $^{40}\mathrm{Ca}^+$ fluorescence spectrum and the resonance frequency of $^{44}\mathrm{Ca}^+$ ions, as shown in Fig.~\ref{fig:Figure4}(a), combined with the high natural abundance of $^{40}\mathrm{Ca}$ (96.9\%).

\par
Based on the selective trapping experiment, it took approximately 30 s to trap $^{44}\mathrm{Ca}^+$ ions. Therefore, in the sympathetic cooling experiment, we kept the oven on for 30 s to load the ions. As a result, $^{44}\mathrm{Ca}^+$ ions were trapped via sympathetic cooling, and a few seconds after turning off the oven, the position fluctuations of $^{40}\mathrm{Ca}^+$ stabilized. This suggests that the cooling time of $^{44}\mathrm{Ca}^+$ ions was on the order of a few seconds, which is one to two orders of magnitude longer than the theoretical estimate based on the numerical calculations by Guggemos \textit{et al}. \cite{Sympathetic_cooling_hot_ion}. The main factors contributing to this discrepancy are likely the nonisotropic nature of the trapping potential and the higher-than-expected initial ion energy. Guggemos \textit{et al}. reported cases where the experimental cooling time was approximately three orders of magnitude longer than the simulation results, concluding that the difference in the initial energy was a significant factor.

\par
Future research directions include evaluating the cooling temperature of the ion trapping via sympathetic cooling. One critical issue is that when attempting to observe $^{44}\mathrm{Ca}^+$, adjusting the cooling laser wavelength causes $^{40}\mathrm{Ca}^+$ ions to escape, preventing the simultaneous observation of both isotope ions. This issue could be resolved by introducing an additional laser capable of exciting and detecting both $^{44}\mathrm{Ca}^+$ and $^{40}\mathrm{Ca}^+$ ions simultaneously. Future prospects include trapping isotope ions other than $^{44}\mathrm{Ca}^+$, designing a trap with an X-junction structure \cite{doi:10.1126/sciadv.1601540}, and implementing low-heating transport methods that combine sympathetic cooling with transport techniques. Unintended hopping within ion chains during quantum operations is a significant challenge \cite{PhysRevA.83.063401}. Addressing this issue will require improvements in trapping potential anisotropy, laser parameters, and other experimental conditions.

\section*{Data availability}
The data that support the findings of this study are available from the corresponding author upon reasonable request.

\bibliography{sample}

\section*{Acknowledgements}

This work was supported by JST Moonshot R \& D (Grant Number JPMJMS2063).

\section*{Author contributions statement}

M.M. contributed equally to the conceptualization and methodology, led the data curation and validation, and took the lead in writing the original draft and in reviewing and editing the manuscript.
K.F. and N.S. fabricated the ion trap used in this research, contributed to writing the section on the shaped through-hole, and participated in the review and editing process.
T.H. contributed equally to software development and was involved in reviewing and editing the manuscript.
K.H. was equally involved in developing the methodology, providing resources, and software development, and also took part in reviewing and editing.
U.T. contributed equally to the conceptualization, methodology, and resource provision, led project administration and supervision, and participated in funding acquisition and manuscript review and editing.

\section*{Additional information}

The authors declare no competing interests.

\end{document}